\documentclass{article}
\usepackage{spconf,amsmath,graphicx}
\usepackage{verbatim}
\usepackage{amsfonts}
\usepackage[inline]{enumitem}
\usepackage{multirow}
\usepackage{balance}
\usepackage{lipsum}
\usepackage{amssymb}
\usepackage{url}

\title{Towards Transferable Speech Emotion Representation: On loss functions for cross-lingual latent representations}
\name{Sneha Das$^1$, Nicole Nadine Lønfeldt$^2$, Anne Katrine Pagsberg$^{2, 3}$, Line H. Clemmensen$^1$
}
\address{$^1$Department of Applied Mathematics and Computer Science, Technical University of Denmark\\
$^2$Child and Adolescent Mental Health Center, Copenhagen University Hospital, Capital Region\\
$^3$Faculty of Health, Department of Clinical Medicine, Copenhagen University}
\begin{document}
%
\maketitle
\begin{abstract}
In recent years, speech emotion recognition~(SER) has been used in wide ranging applications, from healthcare to the commercial sector. In addition to signal processing approaches, methods for SER now also use deep learning techniques which provide transfer learning possibilities. However, generalizing over languages, corpora and recording conditions is still an open challenge. In this work we address this gap by exploring loss functions that aid in transferability, specifically to non-tonal languages. We propose a variational autoencoder (VAE) with KL annealing and a semi-supervised VAE to obtain more consistent latent embedding distributions across data sets. To ensure transferability, the distribution of the latent embedding should be similar across non-tonal languages (data sets). We start by presenting a low-complexity SER based on a denoising-autoencoder, which achieves an unweighted classification accuracy of over 52.09\% for four-class emotion classification. This performance is comparable to that of similar baseline methods. Following this, we employ a VAE, the semi-supervised VAE and the VAE with KL annealing to obtain a more regularized latent space. We show that while the DAE has the highest classification accuracy among the methods, the semi-supervised VAE has a comparable classification accuracy and a more consistent latent embedding distribution over data sets.\footnote{Link to code: \url{https://bit.ly/34CgkSZ}}
\end{abstract}

\section{Introduction}

Speech emotion recognition~(SER) refers to a group of algorithms that deduce the emotional state of an individual from their speech utterances. SER combined with affect recognition using other modalities, like vision and physiological signals, are deployed in a wide range of applications. For instance, in the detection and intervention of disorders in healthcare, monitoring the attentiveness of students in schools, risk assessment within the criminal justice system, and for commercial applications, like detecting customer satisfaction in call-centers and by employment agencies to find suitable candidates.

\begin{figure}[!t]
\centering
\includegraphics[width=0.9\columnwidth]{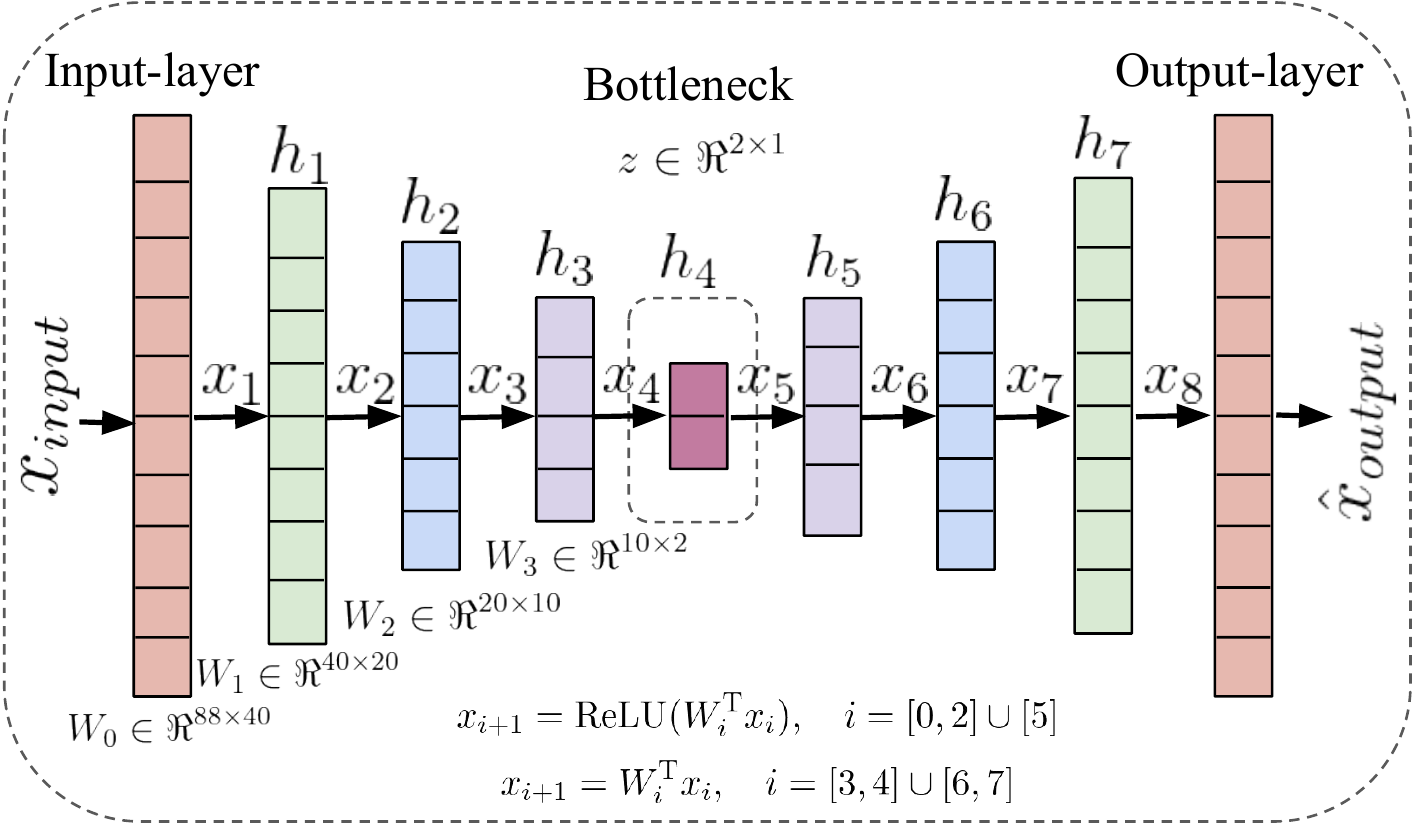}
\caption{Illustration of the architecture employed for all the models explored in this work.}
\vspace{-0.5cm}
\label{fig:archi}
\end{figure}

State-of-the-art SER techniques have evolved from the more conventional signal processing and machine learning based methods to deep neural network based solutions~\cite{akccay2020speech}. Classical methods were based on hidden Markov models~(HMM), Gaussian mixture models~(GMM), support vector machines~(SVM) and decision trees. Contributions based on HMM employed energy and pitch features~\cite{hmm1} showed high classification accuracy. Spectral, prosodic and energy features in tandem with a GMM and SVM were used to recognize emotions from Basque and Chinese datasets~\cite{luengo2005automatic, GMM_supervector}, while pitch based input features were used to obtain a GMM and tested on more heterogeneous speech corpora~\cite{busso_GMM}. SVM is another popular method, either used as the primary classification tool, or in coordination with other techniques to predict the affect classes~\cite{friedman2001elements}. 

Since the successful re-emergence of neural networks, more advanced methods that use deep learning have been used for SER. Long short-term memory~(LSTM), bidirectional LSTM and recurrent neural networks~(RNN) were used to predict the quadrant in the dimensional emotional model~\cite{lstm_rnn_blstm}. Convolutional neural networks, having had immense success in computer vision, are a common architecture choice for neural network based SERs. Furthermore, unsupervision and semi-supervision techniques using autoencoders~(AEs) and its regularized variants were used to learn a lower dimensional latent representation for the emotions, that were then employed at various levels to classify speech into emotional categories~\cite{deng2013sparse, xia2013using, ae_domainAdapt}. 

Latent representation based methods, like the AE, are useful in the modelling of emotions from speech signals, because they can compress the input features to a smaller and ideally more target relevant latent embedding. This can also lead to better knowledge transfer between data sets by transferring generic emotions representations to unseen languages and corpora, hence addressing label shortage and in the process making the models relatively more interpretable. DAE was one of the earliest deep learning based unsupervised learning techniques for SER~\cite{xia2013using}. This was followed by the use of sparse AE for feature transfer~\cite{deng2013sparse} and for SER on spontaneous data set~\cite{dissanayake2020speech}. Furthermore, end-to-end representation learning for affect recognition from speech was proposed and showed performance comparable to existing methods~\cite{ghosh2016representation}. In recent years, techniques like variational and adversarial AEs and adversarial variational Bayes have been exploited to learn the latent representations of speech emotions with input features ranging from the raw signals to hand crafted features~\cite{latif2018variational, parthasarathy2019improving, eskimez2018unsupervised, neumann2019improving}. 

To  successfully model the emotion representations, \begin{enumerate*}\item the latent space should be discriminative of the emotion classes, \item distribution of the latent embedding should be similar over different languages and corpora, such that the latent space is invariant to emotion-irrelevant factors while preserving linguistic and cultural differences in expressing emotions. \end{enumerate*}
Despite the use of AEs for SER in existing literature, few methods provide insights beyond classification accuracy on within and cross-corpus datasets. 
The contribution of this work is to present a method that fares well on both the aforementioned factors. We begin by presenting a low-complexity DAE for SER that achieves a performance similar to existing methods, as listed in Tab.~\ref{tab:baseline}. We show that while the DAE discovers a class-discriminative latent space, the distribution of the latent embedding is considerably different for different languages.  Hence, to obtain a more regularized latent space, we employ a variational autoencoder~(VAE). Since the latent space of a VAE is continuous and probabilistic, this approach is a step further towards a dimensional model of emotions, which proposes that a common process generates all affective states, thereby rendering emotions as a more continuous process~\cite{bakker2014pleasure}. While a vanilla-VAE yields consistent latent distributions over cross-lingual corpora, the latent embedding is not discriminative of the emotion classes due to a dominant KL-loss. Therefore, we address the question: what is the optimal balance between the reconstruction loss and the regularization loss to obtain a class-discriminative and distribution-consistent latent representation. Towards that end, we explore \begin{enumerate*}\item a VAE with KL-loss annealing using cyclic scheduling to improve class discrimination. \item To further improve performance, we employ semi-supervised training of the VAE by incorporating a clustering loss in the learning function. \end{enumerate*}   


\begin{figure}[!htb]
\centering
\includegraphics[width=0.95\columnwidth]{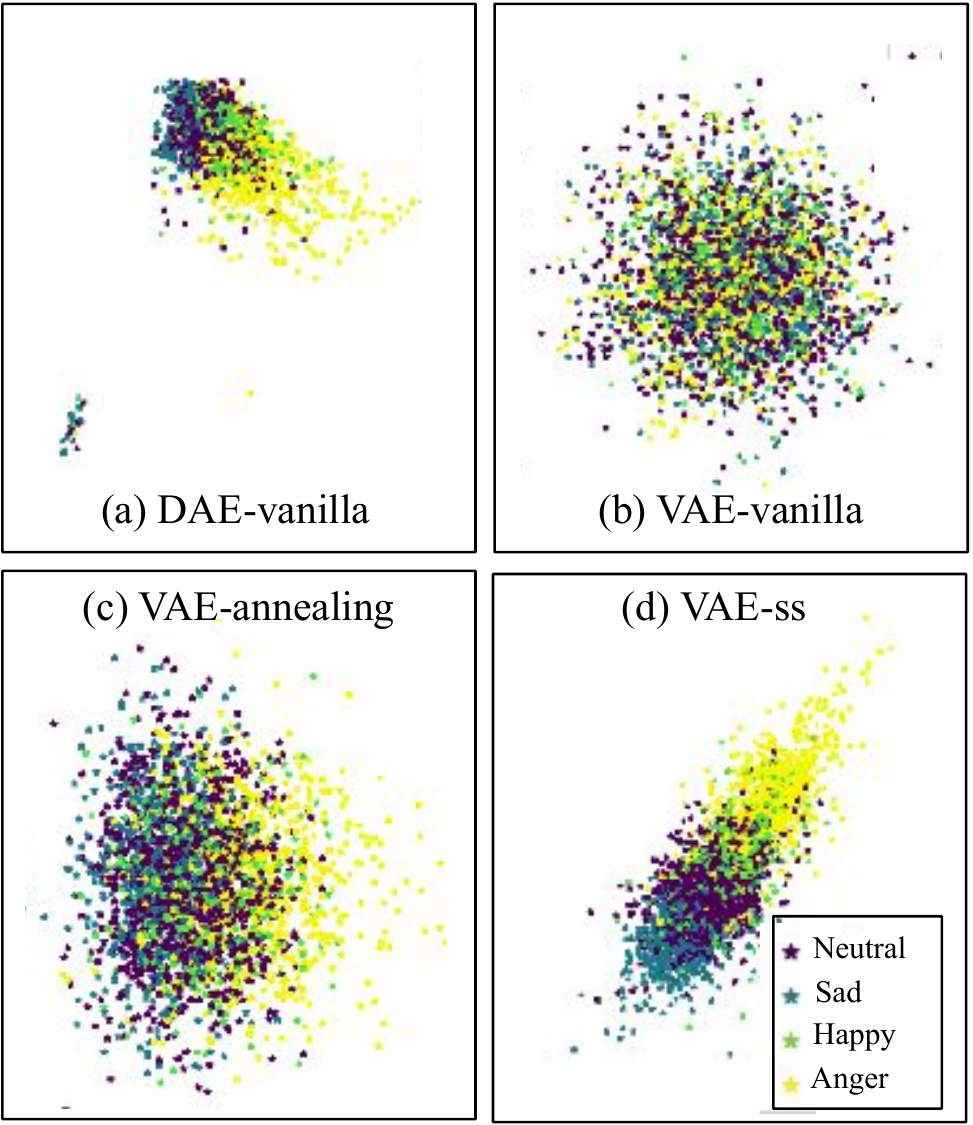}
\caption{Scatterplots of the latent embedding of the training IEMOCAP dataset, obtained from the 4 methods investigated in this paper. The four emotion classes are represented by different colors.}
\label{fig:latentRep}
\vspace{-0.25cm}
\end{figure}

\section{Methodology}
\paragraph*{The standard DAE and VAE:}
\label{sec:1}
A denoising autoencoder can be interpreted as a regularized autoencoder, that directs the network to learn a latent embedding in a noise-free subspace that is representative of the clean input data, from a noisy input. The standard formulation of a DAE loss is:\\ 
\begin{equation}
\begin{aligned}
\text{arg}\min_{f_\theta, g_\phi} \quad \mathcal{L}_{\text{rec}}=\mathbb{E}\lVert \mathbf{x}-g_\phi(f_\theta(\mathbf{x_n})) \rVert_2^2, 
\end{aligned}
\label{eq:1}
\end{equation}
$\mathbf{x_n = x+n}, \quad \mathbf{n}\in\mathcal{N}(0, 1)$, the latent embedding is obtained as $\mathbf{z} = f_\theta(\mathbf{x_n})$ and the reconstructed signal $\mathbf{\hat{x}}=g_\phi(\mathbf{z})$, whereby $f_\theta, g_\phi$ are the encoder and decoder, respectively. 

In a DAE, the latent representation is a point estimate of the training data, whereby the latent space is mostly discrete. According to the dimensional model of emotions and recent methods~\cite{bakker2014pleasure, martinez2014don}, it is more appropriate to model emotions as a more continuous or an ordinal process rather than a categorical process. In order to obtain a more continuous representation of the emotions in the latent space, we employ a VAE. While a VAE is architecturally similar to an autoencoder, the learning goals are considerably different between the two. In other words, in addition to a reconstruction loss, which ensures a latent space representative of the input, a VAE uses the Kullback-Leibler~(KL) loss to minimize the distance between the true and approximated posterior distribution using the KL-divergence~($D_{KL}$). Optimal VAE parameters are obtained by minimizing the following loss: 
\begin{equation}
\begin{aligned}
\text{arg}\min_{\theta, \phi} \quad \mathcal{L}_{\text{rec}}+\mathcal{L}_{\text{KL}}=-\mathbb{E}_{\mathbf{z}\sim q_{\theta}(\mathbf{z|x})}\log p_{\phi}(\mathbf{x|z})\\+D_{KL}(q_{\theta}(\mathbf{z|x})||p(\mathbf{z})), 
\end{aligned}
\label{eq:2}
\end{equation}
where $q_{\theta}(\mathbf{z|x}), p(\mathbf{z})$ are the estimated posterior and the prior distributions, respectively.
\paragraph*{VAE with KL-annealing:}Vanishing KL, also known as posterior collapse occurs when the KL-loss solely directs the minimization of the overall loss. In other words, the reconstruction loss remains consistent whereas the KL-loss reduces incrementally, eventually reducing to a small value. As a consequence, while the posterior distribution resembles the prior distribution with high fidelity~(here a standard Gaussian), z is unrepresentative of data. This can be seen in Fig.~\ref{fig:latentRep}~(b). KL cost annealing with linear or cyclical scheduling is commonly employed to mitigate this issue, wherein a variable weight is added to the KL-loss term~\cite{fu2019cyclical}. We use cyclical annealing in this paper, and the VAE-annealing is expressed as: 

\begin{equation}
\begin{aligned}
\text{arg}\min_{\theta, \phi} \quad \mathcal{L}_{\text{rec}}+\mathcal{L}_{\text{KL}}=-\mathbb{E}_{\mathbf{z}\sim q_{\theta}(\mathbf{z|x})}\log p_{\phi}(\mathbf{x|z})\\+\beta_e D_{KL}(q_{\theta}(\mathbf{z|x})||p(\mathbf{z})), 
\end{aligned}
\label{eq:3}
\end{equation}
where the standard formulation of $\beta_e$:
\begin{equation}
    \beta_{e}= 
\begin{cases}
    f(\tau)=\frac{0.25}{R}\tau, & \tau \leq R\\
    0.25,              & \tau > R \quad \text{where} \quad \tau = \frac{\text{mod}(e-1, \frac{T}{M})}{\frac{T}{M}},
\end{cases}
\label{eq:4}
\end{equation}
$e$ being the training epoch, the total number of cycles given by $M=2$, $T=50$ is the total number of epochs and $R=0.5$ being the proportion of monotonic increase in an annealing cycle.

\paragraph*{VAE with Semi-supervision:}To obtain a latent representation where the emotion classes have improved discrimination while maintaining the regularized space, we incorporate a cluster loss, $\mathcal{L}_{\text{clus}}$ within the VAE. The cluster loss is obtained as the ratio of intra-class and inter-class distances of the latent embedding and minimizing the overall loss will lead to minimize the intra-class distances and maximize the inter-class distance. This is often referred to as the center-loss~\cite{rydhmer2021dynamic}. The overall optimization problem is hence written as: 

\begin{equation}
\begin{aligned}
\text{arg}\min_{\theta, \phi} \quad \mathcal{L}_{\text{rec}}+\beta_e\mathcal{L}_{\text{KL}}+\gamma\mathcal{L}_{\text{clus}},\\
\mathcal{L_{\text{clus}}}= \frac{D_{\text{intra}}}{D_{\text{inter}}} = \frac{\sum\limits_{k=1}^K\sum\limits_{\forall i\in k}D\mathbf{(z_i,\overline{z}^k)}}{\sum\limits_{k=1}^{K-1}\sum\limits_{j=k+1}^{K}D(\mathbf{\overline{z}^k, \overline{z}^j})}, 
\end{aligned}
\label{eq:5}
\end{equation}
where $k$ is the class under consideration, $K$ is the total number of classes, $\overline{z}^k$ is the mean $z$ for class $k$ and $\gamma=0.5$. 

\section{Experiments and Results}

\paragraph*{Datasets and input features:}
We use the IEMOCAP data set, an audio-visual affect data set, to train and validate the models~\cite{busso2008iemocap}. The data set comprises of annotations representing both the categorical and dimensional emotional model~\cite{bakker2014pleasure}. To study how the latent representations are transferred between corpora, we use \begin{enumerate*} \item the Surrey Audio-Visual Expresses Emotion~(SAVEE) database that is primarily English and consists of male speakers only, \item the Berlin Database of Emotional Speech~(Emo-DB) recorded in German and, \item the Canadian French Emotional~(CaFE) speech database comprising of French audio samples~\cite{jackson2014surrey, burkhardt2005database, gournay2018canadian}. \end{enumerate*}, the URDU-dataset consisting of Urdu speech, and the Acted Emotional Speech Dynamic~(AESD) Database comprising of Greek speech~\cite{latif2018cross, vryzas2018speech}. Note that AESD does not comprise of samples from neutral emotion. In this work, we utilize the audio modality only, and train the models with data from the emotional categories {\it neutral~(N), sad~(S), happy~(H), angry~(A)}. We use the extended Geneva minimalistic acoustic parameter set~(eGeMAPS)~\cite{eyben2015geneva}, specifically the functionals of lower-level features. Each speech sample yields a feature vector comprised of 88 features and we use the OpenSmile toolkit to extract the features~\cite{eyben2010opensmile}. 
\vspace{-0.2cm}
\paragraph*{System architecture:}
 While AEs have been employed for SER in existing literature, their main focus was to propose novel network architectures that provide better classification accuracy~\cite{dissanayake2020speech, latif2018variational, neumann2019improving, parthasarathy2019improving, xia2013using}. Since our goal is to explore the cross-lingual transferability of the latent representations, we adhere to the simplest architecture that provides comparable results to previous works.  For a fair comparison between the performance of the models, we design all systems with identical architectures. The input feature dimension is 88 and we maintain the latent dimension size at $z\in\mathrm{R}^{2\times1}$. In relation to past works, the primary source of low-complexity in our proposed architecture is from the latent dimension size. Since the functionals do not have a temporal correlation and the spatial correlation exists between the statistical parameters of a feature only, we use a fully connected neural network~(NN) instead of a recurrent or convolutional type NN. The network architecture is illustrated in Fig.~\ref{fig:archi}.
 
 \vspace{-0.2cm}
 
\paragraph*{Preprocessing:}
Prior to using the data sets for training and testing, we remove the outliers by computing the z-score and eliminating the data samples that have a z-score, $-10>z>10$. We chose a threshold of 10 instead of the standard value of 3 because the goal of this work is to understand the behavior of the models for both typical and atypical rendition of emotions in speech. Therefore, we only remove the extreme outliers. Following this, the data sets are standardized to obtain a normal feature distribution. 
\vspace{-0.2cm}
\paragraph*{Methods and evaluation:}
We train and validate the four models: \begin{enumerate*}\item DAE-vanilla representing a standard denoising autoencoder, \item VAE-vanilla is a standard variational autoencoder, \item VAE-annealing using KL-loss annealing from Eq.~\ref{eq:3} and \item VAE-ss employing semi-supervision from Eq.~\ref{eq:5} \end{enumerate*}, using 5-fold cross-validation on the IEMOCAP database while the transfer data sets are identical over the iterations. The models were trained over 50 epochs and a batch size of 64, and we used the Adam optimizer with the learning rate set to 1e-3. In the following parts, we evaluate the latent embedding by using them as the input features to classify the speech samples into emotional categories using the support vector classifier~(SVC) with a linear kernel. To evaluate the methods, we first present the overall classification accuracy to understand how our presented systems compare to each other. Following that, we compute the similarity of the latent distribution between the training and the cross-lingual transfer data sets. 

\begin{figure}[!tb]
\centering
\includegraphics[width=0.48\columnwidth]{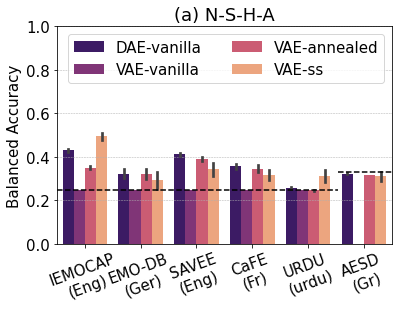}\includegraphics[width=0.48\columnwidth]{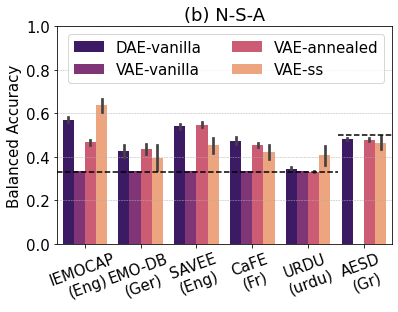}\\
\includegraphics[width=0.48\columnwidth]{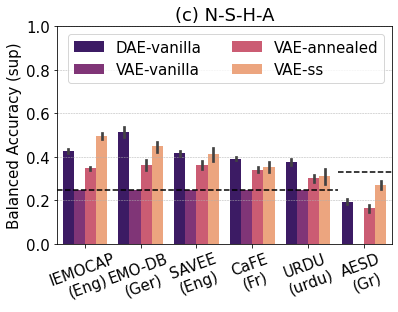}\includegraphics[width=0.48\columnwidth]{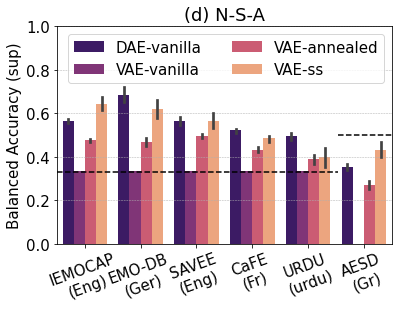}
\caption{(1)~Balanced accuracy on unseen transfer data sets using (a)~4 emotion classes, (b)~3 emotion classes; balanced accuracy with access to 20\% of the unlabeled transfer data sets with (c)~4 emotions and (d)~3 emotion classes.}
\label{fig:acc}
\end{figure}

\subsection{Results}
\begin{table}[!h]

\resizebox{\columnwidth}{!}{%

\begin{tabular}{ |l||l|c|l|  }
 \hline
 {\bf Method} & {\bf Features+Dataset} & {\bf Classes} & {\bf Accuracy}\\
 \hline
 GAN~\cite{latif2019unsupervised}   & eGeMAPS~\cite{eyben2015geneva}+EMO-DB    &2&   66\%~(UAR)\\
FLUDA~\cite{ahn2021cross}&   IS10~\cite{schuller2010interspeech}+IEMOCAP & 4   &50\%~(UA)\\
VAE+LSTM~\cite{latif2018variational}& LogMel+IEMOCAP & 4 & 56.08\%~(UA)\\
AE+LSTM~\cite{latif2018variational}& LogMel+IEMOCAP & 4 & 55.42\%~(UA)\\
Stacked-AE+BLSTM-RNN~\cite{ghosh2016representation}& COVAREP+IEMOCAP~\cite{degottex2014covarep} & 4 & 50.26\%~(UA)\\
DAE+Linear-SVM~\bf(proposed)& eGeMAPS+IEMOCAP & 4 & 52.09\%~(UA)\\
 \hline
\end{tabular}
}
 \caption{Performance of baseline methods in terms of unweighted accuracy~(UA) and unweighted average recall~(UAR)}
\label{tab:baseline}
\vspace{-0.25cm}
 \end{table}
 
 For a fair comparison to existing methods, we list methods based on a similar concept as that in this paper and their performance in Tab.~\ref{tab:baseline}. Classification and consistency results are provided in the following sections. 
\paragraph*{Classification of emotion classes:}

To study the limits of the methods, we evaluate the systems under the following two conditions wherein: 1) the transfer data sets are completely unseen, (2) we have access to 20\% of the unlabeled transfer data sets, whereby we use them to normalize the input feature space. The mean and 95\% CI of the balanced accuracy over 5-folds for conditions 1 and 2 are shown in Fig.~\ref{fig:acc}~(a, b) and Fig.~\ref{fig:acc}~(c, d) respectively. As can be observed from the plot, the DAE-vanilla consistently performs well and is closely followed by VAE with semi-supervision, specifically under condition 2. This indicates that DAE is best at forming a discriminating latent space. We also observe that using a minor portion of the target data set to normalize the input feature space considerably improves the classification accuracy. This is a realistic assumption because open unlabelled language specific data sets can be used for this purpose. Furthermore, VAE-vanilla seems to have no better performance than a random classifier, which is due to posterior collapse as discussed in Sec.~\ref{sec:1}. VAE-annealing seems to perform better under the condition 1 for all transfer data sets, except for the URDU.

On a general level, we observe that the models are able to discriminate between neutral, sad and anger better than between anger and happy. In other words, the models are unable to effectively discriminate between anger and happy, although they represent negative and positive emotions, respectively. This can be further observed in Fig.~\ref{fig:latentRep}~(d), wherein happy and anger samples overlap in the latent space. A probable reason for this could be that the latent representations correspond to activation more than valence.

\paragraph*{Consistency of latent distribution}
\begin{figure}[!tb]
\centering
\includegraphics[width=0.95\columnwidth]{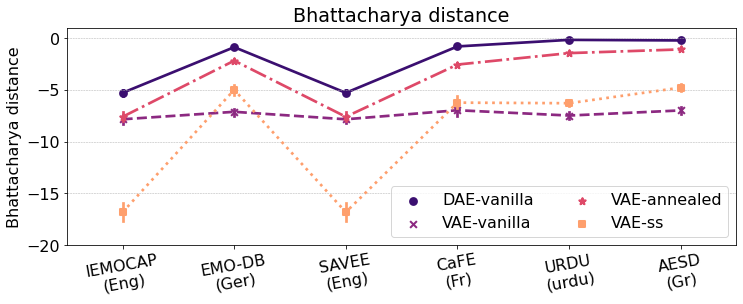}

\caption{Plot presents the mean of the logarithm of the Bhattacharya distance~(BD) of the distribution of the latent embedding from the methods over the transfer datasets with respect to the training data set. Lower BD implies more similarity between distributions.}
\label{fig:BD}
\vspace{-0.25cm}
\end{figure}

\begin{figure}[!tb]
\centering
\includegraphics[width=0.49\columnwidth]{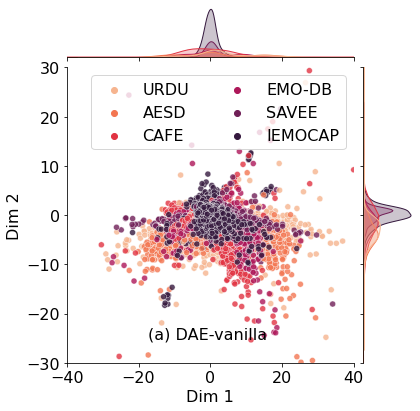}\includegraphics[width=0.49\columnwidth]{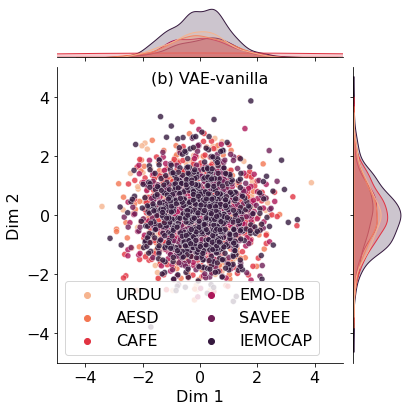}\\
\includegraphics[width=0.49\columnwidth]{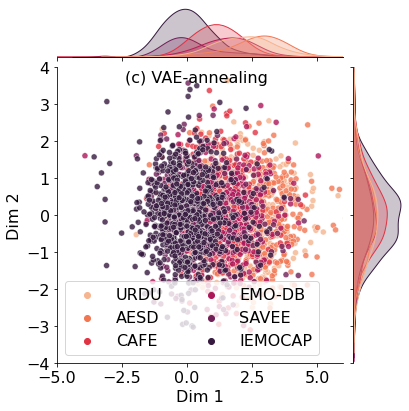}\includegraphics[width=0.49\columnwidth]{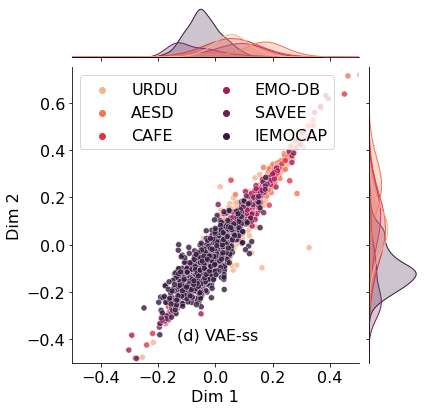}%
\caption{Scatter plots depicting the overlap between the latent embedding obtained from the methods investigated for all the transfer data sets.}
\vspace{-0.05cm}
\label{fig:transfer}
\end{figure}

We use the Bhattacharya distance to quantitatively evaluate the similarity between the latent embedding distribution. The Bhattacharya distance is given as $BD = -\log(\sum\limits_{x\in Z}\sqrt{p_{\text{ref}}(z)p_{\text{i}}(z)})$ between the kernel density estimates~(KDEs) of the reference~(training) dataset and the transfer dataset $i$. We present the logarithm of the BD in Fig.~\ref{fig:BD}; lower BD value implies that the transfer data set is more similar to the reference data set. Furthermore, the scatter plots of the latent embedding for the four methods over the training and transfer data sets are illustrated in Fig.~\ref{fig:transfer}. VAE-vanilla and VAE-ss are shown to have the lowest BD with respect to the training data set and this can also be observed in Fig.~\ref{fig:transfer}~(b, d), where the scatter plots of all the transfer data sets overlap. While VAE-annealing and VAE-ss indicate an offset between the data sets in Fig.~\ref{fig:transfer}~(c, d), the covariance for the data sets remain similar. In contrast, for DAE-vanilla, the distribution of the latent embedding have a relatively higher BD and low overlap, as observed in Figs.~\ref{fig:BD}~\&~\ref{fig:transfer}. 

\section{Conclusions}
In this work, we explored latent representation based methods for their ability to transfer emotion representations across cross-lingual data sets for non-tonal languages. Our objective was to find representations motivated from the dimensional emotion model, wherein emotions are processed on the relative and continuous scale. We evaluated the methods on class separability in the latent space and the consistency of the latent distribution over the data sets. We present a VAE with KL annealing and semi supervision incorporating cluster distance within the loss.  The standard DAE yielded the latent embedding with highest classification accuracy implying high class discrimination. However, VAE with semi-supervision not only provided a comparable classification accuracy to the DAE, but also a more consistent latent space over the transfer datasets. Furthermore, a natural progression of this work would be to employ semi-supervision using a continuous distance metric instead of a classification loss and is left for future work.

\bibliographystyle{IEEEbib}
\bibliography{refs}
\balance
\label{sec:refs}
\end{document}